# Influence of intrinsic decoherence on entanglement teleportation via a Heisenberg XYZ model with different Dzyaloshinskii-Moriya interaction

**Meng Qin · Zhong-Zhou Ren**

**Abstract** We investigate the characteristics of entanglement teleportation of a two-qubit Heisenberg XYZ model under different Dzyaloshinskii-Moriya interaction with intrinsic decoherence taken into account. The comparison of the two different Dzyaloshinskii-Moriya interaction, the effects of the initial state and the inputting state on the entanglement teleportation are presented. The results reveal that the dynamics of entanglement is a symmetry function about $J$ for the $D_z$ system, whereas it is not for the $D_x$ system. The ferromagnetic case is superior to the antiferromagnetic case for restrain decoherence when using the $D_x$ system. The dependence of entanglement, output entanglement, fidelity on initial state angle $\alpha$ all demonstrate periodic. Moreover, we find that seemingly some system are not suitable for teleportation, but they can acquire some best exhibition if we take the proper initial state and inputting state.

## 1 Introduction

Quantum entanglement is the fundamental characteristic of quantum mechanics, and it is also an important resource for quantum communication and quantum computation [1].

M. Qin (✉) · Z. -Z. Ren

Department of Physics, Nanjing University, Nanjing 210093, People's Republic of China

e-mail: qrainm@gmail.com

M. Qin

College of Sciences, PLA University of Science and Technology, Nanjing 211101, People's Republic of China

Z. -Z. Ren

Center of Theoretical Nuclear Physics, National Laboratory of Heavy-Ion Accelerator, Lanzhou 730000, People's Republic of China

Kavli Institute for Theoretical Physics, Chinese Academy of Sciences, Beijing 100190, People's Republic of China

Recently, much attention has been paid to the entanglement in spin model [2,3] since its play a key role in quantum information processing task [4,5]. In particular, the properties of Heisenberg model with Dzyaloshinski-Moriya (DM) exchange interaction [6,7] (arising from spin-orbit coupling) have been studied extensively for this interaction causes another type of anisotropy. Many physical systems, such as quantum dots, nuclear spins, optical lattices, and superconductors also have been simulated by the model[8]. It is now well established that at low temperatures these systems exhibit new types of magnetic order and novel quantum phases[9]. Recently, Li et al [6] discuss the Heisenberg model with different DM interaction. They find that the enhancement entanglement and the higher critical temperature can be obtained by adjusting the values of different DM interaction. This means that a more efficient control parameter can be obtained by adjusting the direction of DM interaction, no matter in the antiferromagnetic case or in the ferromagnetic case. Induced by this, if we use the Heisenberg XYZ model with different DM interaction to perform quantum teleportation protocols, the entanglement of replica states and the fidelity of the teleportation will also have another means to manipulate.

As far as we are aware, the entanglement in the system must be maintained a long time in order to fulfill the quantum task. The unavoidable interaction of a system with its surroundings always make entanglement decay with time. It's difficult to keep the coherence of a quantum state as the quantum system correlate with it external environment. This will induce entanglement of sudden death and so on. So, all the discussions must be including the environmental effect on the system[10,11]. In recent years, there have been many proposals to solve the decoherence problem which is responsible for quantum to classical transition. The general investigation method for this problem is tracing out all other degrees except quantum states of interest. However, Milburn[12] has given a simple model of intrinsic decoherence based on an assumption that for sufficiently short time steps the system does not evolve continuously under unitary evolution but rather in a stochastic sequence of identical unitary transformations. This assumption leads to a simple modification of standard quantum mechanics and the quantum coherence is automatically destroyed as the quantum system evolves [13,14].

The model gives a valuable explanation to the disappearance of quantum state coherence and it's a meaningful application of quantum mechanics in macroscopic system. In particularly, this model is convenient to get exact analytical solutions and its stochastic behavior in time evolution can be an effective approximation for describing the phenomenon of the system [15].

The effects of intrinsic decoherence on the dynamics of entanglement have been studied in a number of works. For example, Hu [15] has shown that the ideal spin channels will be destroyed by the intrinsic decoherence environment. Yu [16] has discussed the intrinsic decoherence effects on the entanglement of a two-qubit anisotropic Heisenberg XYZ model. Also, the results of reference [17] show that for an initial pure state of the resource, which is the entangled state of a two-qubit XXZ Heisenberg chain, an inhomogeneous magnetic field can reduce the effects of intrinsic decoherence. Fan has found that the phase decoherence rate makes the original harmonic vibration with respective to time decay to a stable value[18]. All those works enrich our understanding of the decoherence mechanism. However, how the different DM interaction affect the dynamics of entanglement teleportation and the fidelity of entanglement teleportation in Heisenberg XYZ model has not been reported.

In this paper, we investigate the influence of the intrinsic decoherence on the entanglement and entanglement teleportation of a two qubit Heisenberg XYZ model under the influence of different DM interaction. The outline of this work is as follows. In section 2 we introduce the Hamiltonian of the Heisenberg model with different DM interaction and present the exact solution of the model. In section 3, we discuss the evolution of entanglement with different DM interaction. In section 4, entanglement teleportation processes via the above system are investigated. In section 5, we study the effect of initial state and inputting state on the fidelity. Finally, in section 6, we summarize our results and draw our conclusions.

## 2 Model and solution

The Hamiltonian $H_z$ for a two-qubit anisotropic Heisenberg XYZ model with $z$-component DM interaction parameter $D_z$ is

$$H_z = J(1+\gamma)\sigma_1^x\sigma_2^x + J(1-\gamma)\sigma_1^y\sigma_2^y + J_z\sigma_1^z\sigma_2^z + D_z\left(\sigma_1^x\sigma_2^y - \sigma_1^y\sigma_2^x\right) \tag{1}$$

where $J$ and $J_z$ are the real coupling coefficients, $\gamma$ is the anisotropic parameter. $D_z$ is the $z$-component DM interaction parameter, and $\sigma^i$ ($i = x, y, z$) are Pauli matrices. All the parameters are dimensionless.

Obviously, the eigenvalues and eigenvectors of the Hamiltonian $H_z$ are given by

$$\begin{aligned}
E_{z1} &= J_z + 2J\gamma \\
E_{z2} &= J_z - 2J\gamma \\
E_{z3} &= -J_z + 2\sqrt{J^2 + D_z^2} \\
E_{z4} &= -J_z - 2\sqrt{J^2 + D_z^2} \\
|\psi_{z1,z2}\rangle &= \frac{|00\rangle \pm |11\rangle}{\sqrt{2}} \\
|\psi_{z3,z4}\rangle &= \frac{|01\rangle \pm \chi|10\rangle}{\sqrt{2}}
\end{aligned} \tag{2}$$

where $\chi = \dfrac{J - iD_z}{\sqrt{J^2 + D_z^2}}$

The Hamiltonian $H_x$ for a two-qubit anisotropic Heisenberg XYZ model with $x$-component DM interaction parameter $D_x$ is

$$H_x = J(1+\gamma)\sigma_1^x\sigma_2^x + J(1-\gamma)\sigma_1^y\sigma_2^y + J_z\sigma_1^z\sigma_2^z + D_x\left(\sigma_1^y\sigma_2^z - \sigma_1^z\sigma_2^y\right) \tag{3}$$

The eigenvalues and eigenvectors of the Hamiltonian $H_x$ are given by

$$\begin{aligned}
E_{x1} &= 2J - J_z \\
E_{x2} &= 2J\gamma + J_z \\
E_{x3} &= -J(1+\gamma) + \sqrt{[J(1-\gamma)+J_z]^2 + 4D_x^2} \\
E_{x4} &= -J(1+\gamma) - \sqrt{[J(1-\gamma)+J_z]^2 + 4D_x^2} \\
|\psi_{x1}\rangle &= \frac{|01\rangle + |10\rangle}{\sqrt{2}} \\
|\psi_{x2}\rangle &= \frac{|00\rangle + |11\rangle}{\sqrt{2}} \\
|\psi_{x3}\rangle &= \frac{\sin\varphi_1|00\rangle - i\cos\varphi_1|01\rangle + i\cos\varphi_1|10\rangle - \sin\varphi_1|11\rangle}{\sqrt{2}} \\
|\psi_{x4}\rangle &= \frac{\sin\varphi_2|00\rangle + i\cos\varphi_2|01\rangle - i\cos\varphi_2|10\rangle - \sin\varphi_2|11\rangle}{\sqrt{2}}
\end{aligned} \tag{4}$$

here $\varphi_{1,2} = \arctan\left(\dfrac{2D_x}{\sqrt{[J(1-\gamma)+J_z]^2 + 4D_x^2} \mp J(1-\gamma) \pm J_z}\right)$

The master equation describing the intrinsic decoherence under the Markovian approximations is given by[12,14]

$$\frac{d\rho(t)}{dt} = -i[H,\rho(t)] - \frac{\Gamma}{2}[H,[H,\rho(t)]], \qquad (5)$$

where $\Gamma$ is the intrinsic decoherence rate. The formal solution of the above master equation can be expressed as

$$\rho(t) = \sum_{k=0}^{\infty} \frac{(\Gamma t)^k}{k!} M^k \rho(0) M^{+k} \qquad (6)$$

where $\rho(0)$ is the density operator of the initial system and $M^k$ is defined by

$$M^k = H^k e^{-iHt} e^{-\frac{\Gamma t}{2} H^2} \qquad (7)$$

According to equation (6) it is easy to show that, under intrinsic decoherence, the dynamics of the density operator $\rho(t)$ for the above mentioned system which is initially in the state $\rho(0)$ is given by

$$\rho(t) = \sum_{mn} \exp\left[-\frac{\Gamma t}{2}(E_m - E_n)^2 - i(E_m - E_n)t\right] \times \langle\psi_m|\rho(0)|\psi_n\rangle|\psi_m\rangle\langle\psi_n| \qquad (8)$$

where $E_m, E_n$ and $|\psi_m\rangle, |\psi_n\rangle$ are the eigenvalues and the corresponding eigenvectors of $H_{z/x}$ given in equations (2) and (4).

One considers the initial state $|\varphi(0)\rangle = \cos\alpha|01\rangle + \sin\alpha|10\rangle$, the density matrices with time evolution for $D_z$ system in the standard basis $(|00\rangle, |01\rangle, |10\rangle, |11\rangle)$ will be:

$$\rho(t) = \begin{bmatrix} \rho_{11}(t) & \rho_{12}(t) & \rho_{13}(t) & 0 \\ \rho_{21}(t) & \rho_{22}(t) & \rho_{23}(t) & 0 \\ \rho_{31}(t) & \rho_{32}(t) & \rho_{33}(t) & 0 \\ 0 & 0 & 0 & 0 \end{bmatrix} \qquad (9)$$

where $\rho_{11}(t) = \dfrac{\chi^2}{4}\cos^2\alpha$ , $\rho_{22}(t) = \dfrac{\chi^2}{4}(\cos\alpha + \chi\sin\alpha)^2$ ,

$$\rho_{33}(t) = \frac{1}{4}\left[\cos\alpha(\cos\alpha + \chi\sin\alpha)(2+\eta+\xi) + \chi^2\sin^2\alpha\right],$$

$$\rho_{21}(t) = -\frac{\chi^2\eta}{4}\cos\alpha(\cos\alpha + \chi\sin\alpha), \quad \rho_{31}(t) = -\frac{\chi\cos\alpha}{4}(\cos\alpha + \eta\chi\sin\alpha + \eta\cos\alpha),$$

$$\rho_{12}(t) = -\frac{\xi\chi^2\cos\alpha}{4}(\cos\alpha + \chi\sin\alpha),$$

$$\rho_{13}(t) = -\frac{\chi\cos\alpha}{4}(\cos\alpha + \xi\chi\sin\alpha + \xi\cos\alpha),$$

$$\rho_{32}(t) = \frac{\chi}{4}(\cos\alpha + \chi\sin\alpha)(\cos\alpha + \chi\sin\alpha + \xi\cos\alpha),$$

$$\rho_{23}(t) = \frac{\chi}{4}(\cos\alpha + \chi\sin\alpha)(\cos\alpha + \chi\sin\alpha + \eta\cos\alpha), \quad \eta = e^{-t(E_3-E_4)(\Gamma E_3 - \Gamma E_4 + 2i)/2},$$

$$\xi = e^{-t(E_3-E_4)(\Gamma E_3 - \Gamma E_4 - 2i)/2}$$

The density matrix of the $D_x$ system is too complicated to write out here, so we will mainly discuss the numerical results for the $D_x$ case.

### 3 Entanglement evolutions

To quantify the amount of entanglement associated with $\rho(t)$, we consider the concurrence, which is defined as[19,20].

$$C = \max\left[0, 2\max(\lambda_i) - \sum_i^4 \lambda_i\right] \qquad (10)$$

where $\lambda_i$ are the square roots of the eigenvalues of the matrix $R = \rho(t)S\rho^*(t)S$, $\rho(t)$ is the density matrix, $S = \sigma_1^y \otimes \sigma_2^y$, and the asterisk stands for the complex conjugate. After some straightforward algebra, we can get the eigenvalues of $D_z$ system.

$$\lambda_1 = \lambda_2 = 0$$

$$\lambda_3 = \sqrt{|\rho_{22}(t)\rho_{33}(t) + \rho_{23}(t)\rho_{32}(t) + 2\sqrt{\rho_{23}(t)\rho_{32}(t)\rho_{22}(t)\rho_{33}(t)}|} \qquad (11)$$

$$\lambda_4 = \sqrt{|\rho_{22}(t)\rho_{33}(t) + \rho_{23}(t)\rho_{32}(t) - 2\sqrt{\rho_{23}(t)\rho_{32}(t)\rho_{22}(t)\rho_{33}(t)}|}$$

So the analytical result of concurrence for $D_z$ system can be got by taking Eq. (11) into Eq. (10). The concurrence of $D_x$ interaction is also a function of the initial state and the system parameters. In the following, we will study their behaviors.

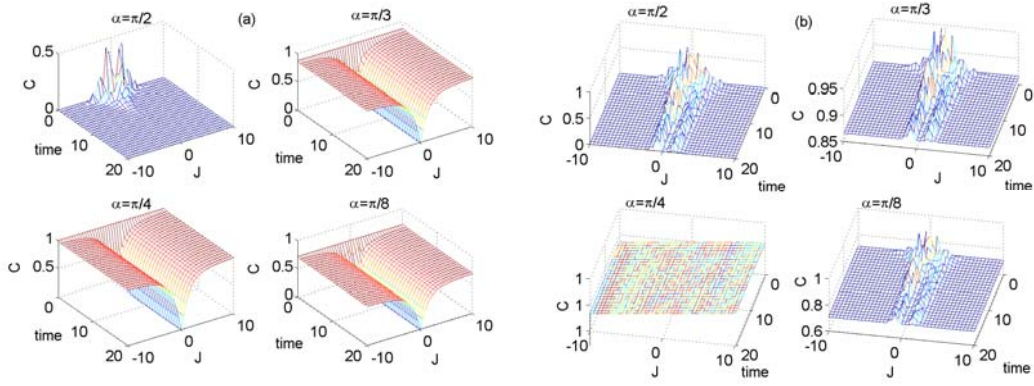

**Fig. 1** Concurrence versus the real coupling coefficients $J$ and the time $t$ for different initial state when system in (a) Eq.(1), (b) Eq.(3). The parameter values for the plot are $\gamma=0.2$, $J_z=1$, $D_{z/x}=2$, $\Gamma=0.02$.

In Fig.1 we depict the concurrence as a function of the real coupling coefficients $J$ and the time $t$ for different initial condition. As is shown in Fig. 1a, the concurrence is a symmetry function about $J=0$ regardless of the value of $\alpha$ for $D_z$ system. The concurrence will decrease with the increase of $|J|$ and the time $t$ when the initial state is $|10\rangle$. For $\alpha=\dfrac{\pi}{3},\dfrac{\pi}{4}$, and $\dfrac{\pi}{8}$ the concurrence holds a stationary value with the deocoherence and it's an increasing function about $|J|$. Clearly, the entanglement can attain the max value 1 when initial state is $|\varphi(0)\rangle=\dfrac{1}{\sqrt{2}}(|01\rangle+|10\rangle)$. So the Bell state is the optimal initial state here. In Fig. 1b, we see that the concurrence is no longer a symmetry function with the increase of $|J|$ except the $\alpha=\dfrac{\pi}{4}$ case for the $D_x$ system. So the evolution of entanglement is different for the antiferromagnetic case and the ferromagnetic case. Obviously, the entanglement in ferromagnetic case is stronger than that in the antiferromagnetic case. This phenomenon shows that the ferromagnetic model is more superior to the antiferromagnetic model for restrain decoherence when using $D_x$ system. The Bell state is still the best initial state in this condition, except that there is other optimal state which can be used as initial state, i.e. the concurrence can keep a higher value when we choose the initial state as $|\varphi(0)\rangle=\dfrac{1}{2}|01\rangle+\dfrac{\sqrt{3}}{2}|10\rangle$.

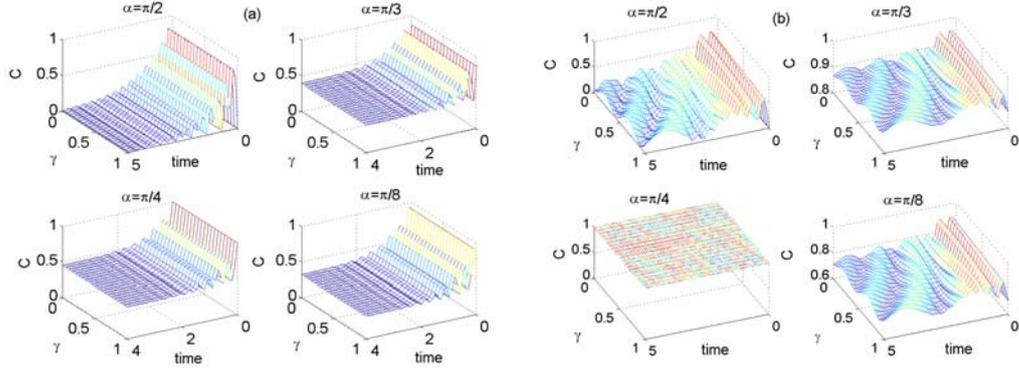

**Fig. 2** Concurrence versus the anisotropic parameter $\gamma$ and the time $t$ for different initial state when system in (a) Eq.(1), (b) Eq.(3). The parameter values for the plot are $J=1$, $J_z=1$, $D=2$, $\Gamma=0.02$.

In Fig. 2, the concurrence as a function of the anisotropic parameter $\gamma$ and the time $t$ is plotted. Fig. 2a shows that the entanglement will decrease with the increase of time but keep a stable value with the change of $\gamma$ for every initial state. After the max decoherence time, the concurrence will stay steady value for different initial state angle. The stationary entanglement value is 0, 0.387, 0.447, and 0.316 accordingly. Fig. 2b exhibits the properties of $D_x$ system. The entanglement behaves as a wave function as the time evolves and anisotropic parameter $\gamma$ increases. But the condition is different when the initial state is Bell-state $|\varphi(0)\rangle = \frac{1}{\sqrt{2}}(|01\rangle+|10\rangle)$, the entanglement will keep max value 1.

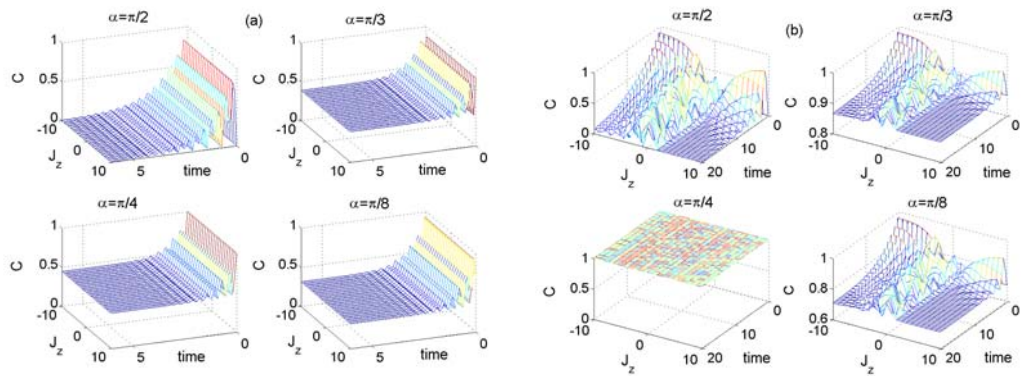

**Fig. 3** Concurrence versus the real coupling coefficients $J_z$ and the time $t$ for different initial state when system in (a) Eq.(1), (b) Eq.(3). The parameter values for the plot are $J=1$, $\gamma=0.5$, $D_{z/x}=2$, $\Gamma=0.02$.

In Fig. 3, the entanglement is plotted versus the real coupling coefficients $J_z$ and the time $t$. The variation of entanglement versus anisotropy coupling coefficients $J_z$ and time $t$ is similar with Fig. 2a. We see that the increase of time can make entanglement decrease, and the entanglement is also a symmetry function about $J_z$. If the system in Eq. (3), as is shown in Fig. 3b, the entanglement is asymmetric function with respect to $J_z=0$ except for $\alpha=\frac{\pi}{4}$ (the entanglement is 1 in this case). We can note that the entanglement behaves as a non-monotonic decreasing function for the $D_x$ system. This is in accordance with the previous result. The ferromagnetic model can get enhancement entanglement and restrain decoherence while the antiferromatnetic model cannot.

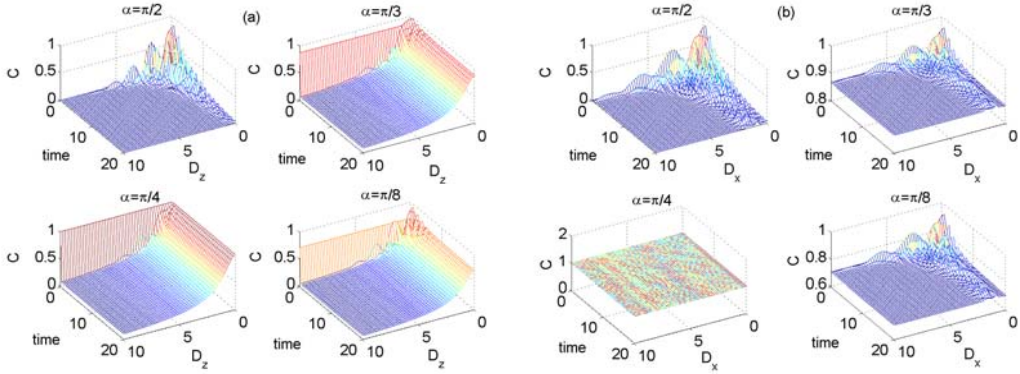

**Fig. 4** Concurrence versus the DM interaction $D_{z/x}$ and the time $t$ for different initial state when system in (a) Eq.(1), (b) Eq.(3). The parameter values for the plot are $J=1$, $\gamma=0.6$, $J_z=1.5$, $\Gamma=0.02$.

In Fig. 4, we give the plot of the entanglement as a function of the DM interaction $D_{z/x}$ and the time $t$ for different values of $\alpha$. Except the initial state $|10\rangle$, the other figures in Fig. 4a show that the entanglement behaves as a monotonic function of DM interaction $D_z$. For a fixed $D_z$, the entanglement holds a stable value with the increase of time. So the initial state $|10\rangle$ is the peculiar state in the $D_z$ system as the concurrence demonstrate different properties. The peculiar state is $|\varphi(0)\rangle = \frac{1}{\sqrt{2}}(|01\rangle+|10\rangle)$ when it comes to $D_x$ system. The entanglement value holds maxmum 1 for this initial state. The entanglement will increase and then decrease with the increase of DM interaction $D_x$ for the other three pictures in Fig. 4b. The entanglement will keep a stable value for the larger $D_x$ value. From the two figures, we notice that the entanglement quickly decreases to 0 when the initial state is $|10\rangle$. This implies that the entanglement state is more suitable than un-entangled state to be chosen as initial state.

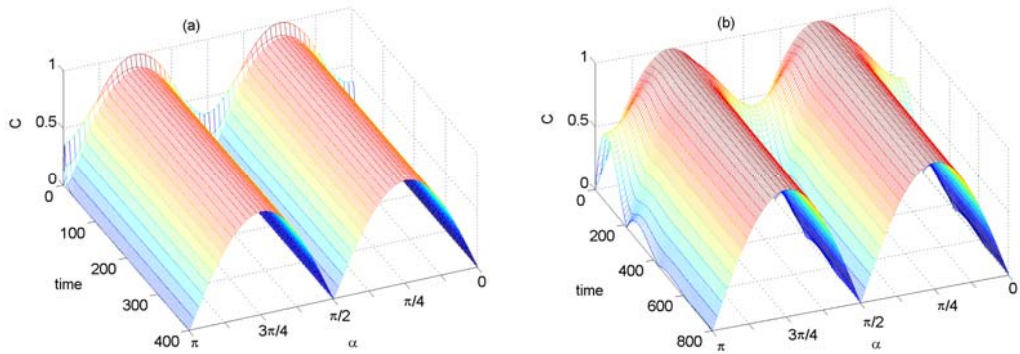

**Fig. 5** Concurrence versus the initial state angle $\alpha$ and the time $t$ for the z-component system when the initial state in (a) $|\varphi(0)\rangle = \cos\alpha|01\rangle+\sin\alpha|10\rangle$, (b) $|\varphi(0)\rangle = \cos\alpha|00\rangle+\sin\alpha|11\rangle$. The parameter values for the plot are $J=1$, $\gamma=0.2$, $J_z=2$, $D_z=0.5$, $\Gamma=0.02$.

Furthermore, in order to know how the initial state affects the entanglement, we plot the concurrence as a function of the initial state angle $\alpha$ and the time $t$ for two kinds of initial condition. Fig. 5 gives the results for the system in z-component DM interaction. In Fig. 5a, we select $|\varphi(0)\rangle = \cos\alpha|01\rangle+\sin\alpha|10\rangle$ as the initial state. The maximal value of concurrence occurs at $\alpha = \frac{\pi}{4}$ and $\alpha = \frac{3\pi}{4}$ in this case. For some initial sate angle $\alpha$, i.e. $\alpha = \frac{\pi}{6}$ or others, the entanglement will show stable characteristic with the decoherence. In Fig. 5b, we change the initial state as $|\varphi(0)\rangle = \cos\alpha|00\rangle + \sin\alpha|11\rangle$. Different with the Fig. 5a, the entanglement will display increase and decrease with the time for some angle $\alpha$, but the max entanglement is still in $\alpha = \frac{\pi}{4}$ and $\alpha = \frac{3\pi}{4}$ point. It is clearly that there exists a periodicity for the initial state angle $\alpha$. The cycle is $\frac{\pi}{2}$ for the two cases. Moreover, we find that there is no difference for the antiferromagnetic and ferromagnetic case in $D_z$ system. The properties are identical if we change the signs of $J$ and $J_z$. This is consistent with the previous conclusions.

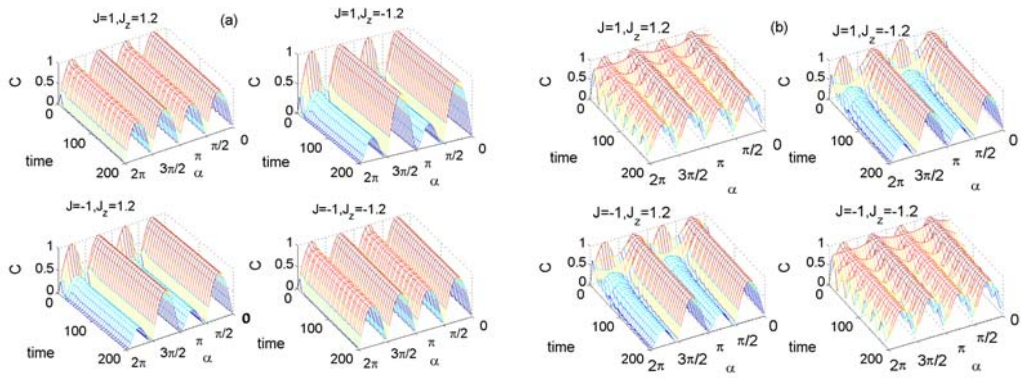

**Fig. 6** Concurrence versus the initial state angle $\alpha$ and the time $t$ for the x-component system when the initial state in (a) $|\varphi(0)\rangle = \cos\alpha|01\rangle + \sin\alpha|10\rangle$, (b) $|\varphi(0)\rangle = \cos\alpha|00\rangle + \sin\alpha|11\rangle$. The parameter values for the plot are $\gamma = 0.2$, $D_x = 0.5$, $\Gamma = 0.02$.

Figure 6 gives the results for the $D_x$ system. We plot the entanglement as a function of the initial state angle $\alpha$ and the time $t$ for different values of $J$ and $J_z$. In Fig. 6a, 6b, we choose the initial state as $|\varphi(0)\rangle = \cos\alpha|01\rangle + \sin\alpha|10\rangle$ and $|\varphi(0)\rangle = \cos\alpha|00\rangle + \sin\alpha|11\rangle$ respectively. We find that the region of entanglement will be greatly affected as long as we alter the sign for one of $J$ or $J_z$, but the results will not be affected if we change their sign together. There is a cycle for the initial state angle $\alpha$ too. The cycle is $\frac{\pi}{2}$ for the same signs of $J$ and $J_z$, but the cycle is $\pi$ for different signs of $J$ and $J_z$.

**4 The effect of initial state and inputting state on entanglement teleportation**

Now we consider Lee and Kim's[21] teleportation protocol using two copies of the above state $\rho(t)\otimes\rho(t')$ as resource. We consider inputting a two-qubit state $|\psi_{in}>=\cos(\theta/2)|10>+e^{i\phi}\sin(\theta/2)|01> \left(0\leq\theta\leq\pi, 0\leq\phi\leq 2\pi\right)$. The output replica state can be obtained by applying a joint measurement and local unitary transformation on the input state. Thus the output state is given by

$$\rho_{out}(t)=\sum_{i,j}p_{ij}(\sigma_i\otimes\sigma_j)\rho_{in}(\sigma_i\otimes\sigma_j) \tag{12}$$

where $\sigma_i (i=0,x,y,z)$ signify the unit matrix $I$ and three components of the Pauli matrix respectively, $\rho_{in}=|\psi>_{in}<\psi|$ and $p_{ij}=tr[E^i\rho(T)]tr[E^j\rho(T)]$, $\sum p_{ij}=1$. Here $E^0=|\psi_{Bell}^2><\psi_{Bell}^2|$, $E^1=|\psi_{Bell}^3><\psi_{Bell}^3|$, $E^2=|\psi_{Bell}^0><\psi_{Bell}^0|$, $E^3=|\psi_{Bell}^1><\psi_{Bell}^1|$ and $|\psi_{Bell}^{0,3}>=\frac{1}{\sqrt{2}}(|00>\pm|11>), |\psi_{Bell}^{1,2}>=\frac{1}{\sqrt{2}}(|01>\pm|10>)$. So the output entanglement $C_{out}=C(\rho_{out})=\max\left[0, 2\max(\lambda_i')-\sum_i^4\lambda_i'\right]$ can be computed out when the parameters of the channel are given.

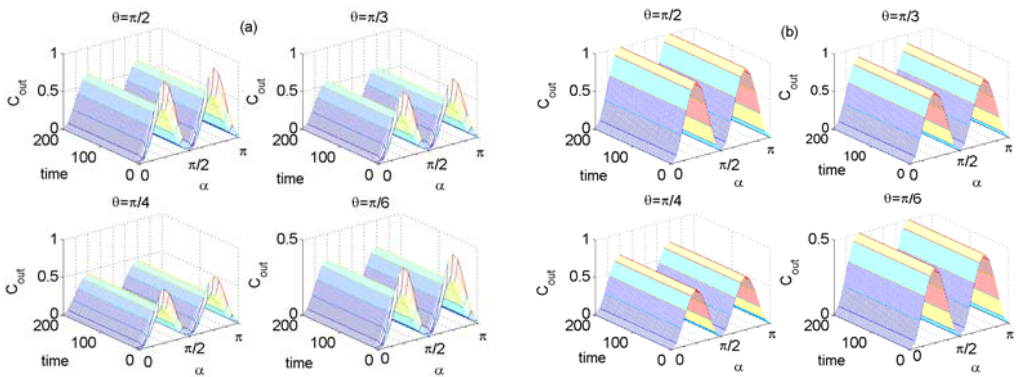

**Fig. 7** Entanglement of output state versus the initial state angle $\alpha$ and the time $t$ for z-component system when the initial state in (a) $|\varphi(0)\rangle=\cos\alpha|01\rangle+\sin\alpha|10\rangle$,

(b) $|\varphi(0)\rangle = \cos\alpha |00\rangle + \sin\alpha |11\rangle$. The parameter values for the plot are $J=1, \gamma=0.2$, $J_z = 2$, $D_z = 0.5$, $\phi = 0$, $\Gamma = 0.02$.

In this section, we start to discuss the properties of output entanglement. In Fig. 7, the output entanglement $C_{out}$ as a function of the initial state angle $\alpha$ and the time is plotted with different inputting state for $D_z$ system. From the two figures, we find that the output entanglement is mainly decreasing with the decrease of inputting state $\theta$. The periodic dependence of output entanglement on the angle $\alpha$ also exists in the figures and the cycle is $\pi/2$. The inputting state will not affect the periodicity of the system. One must notice that the Fig. 7a and the Fig. 7b denote two kinds of initial state. But their cycle is identical here. So we can infer that the z-component DM system-self has more influences on the periodic. In addition, there are some differences between the two figures. i.e. the dependence of output entanglement on time; the maximum value. If we change the sign of $J$ and $J_z$, we still get the same result here. This indicates that the output entanglement for anti-ferromagnetic case and for ferromagnetic case have the same properties in the $D_z$ system.

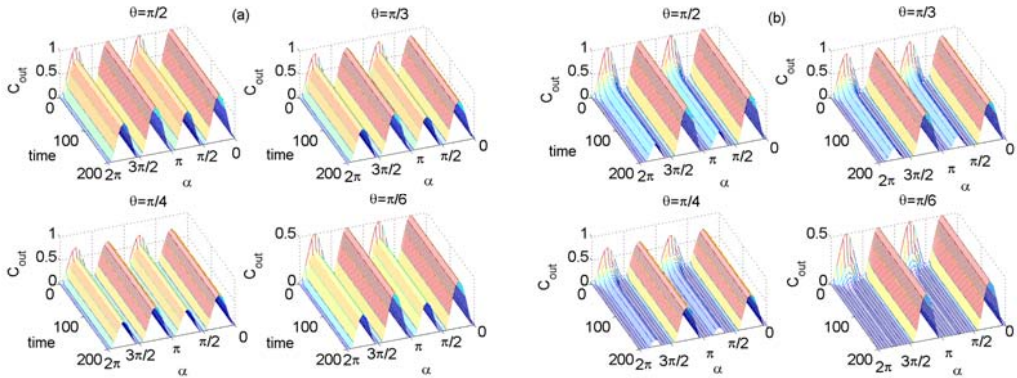

**Fig. 8** Entanglement of output state versus the initial state angle $\alpha$ and the time $t$ for x-component system when the initial state in $|\varphi(0)\rangle = \cos\alpha |01\rangle + \sin\alpha |10\rangle$, The parameter values for the plot are $J=1$, $\gamma = 0.2$, $D_x = 0.5$, $\Gamma = 0.02$. $\phi = 0$ (a) $J_z = 2$ (b) $J_z = -2$.

In Fig. 8, the output entanglement $C_{out}$ as a function of the initial state angle $\alpha$ and the time $t$ is plotted for $D_x$ system. Just like Fig. 7, the output entanglement will also decrease with the decrease of inputting state $\theta$ for the $D_x$ system. The periodicity is $\pi$ in this case. We find the behaviors of the output entanglement is different for the initial angle $\alpha \in [0, \pi/2]$ and $\alpha \in [\pi/2, \pi]$. The change of the system from antiferromagnetic case to ferromagnetic case will greatly reduce the entanglement for the region of $\alpha \in [\pi/2, \pi]$.

**5 The effect of initial state and inputting state on the fidelity**

The fidelity between $\rho_{in}$ and $\rho_{out}$ characterizes the quality of the teleported state $\rho_{out}$. When the input is a pure state, we can apply the concept of fidelity as a useful indicator of teleportation performance of a quantum channel. If the quantum channel is maximal, the best entanglement teleportation will be obtained. The fidelity of $\rho_{in}$ and $\rho_{out}$ is defined to be[22]

$$F(\rho_{in}, \rho_{out}) = \{tr[\sqrt{(\rho_{in})^{1/2} \rho_{out} (\rho_{in})^{1/2}}]\}^2 \tag{13}$$

The two states are equal for $F = 1$ and orthogonal for $F = 0$. Thus, the larger $F$ is, the better the teleportation is achieved[23].

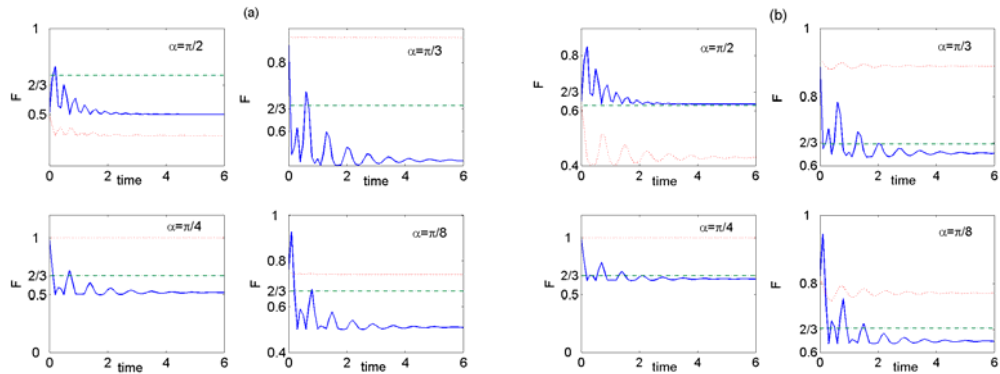

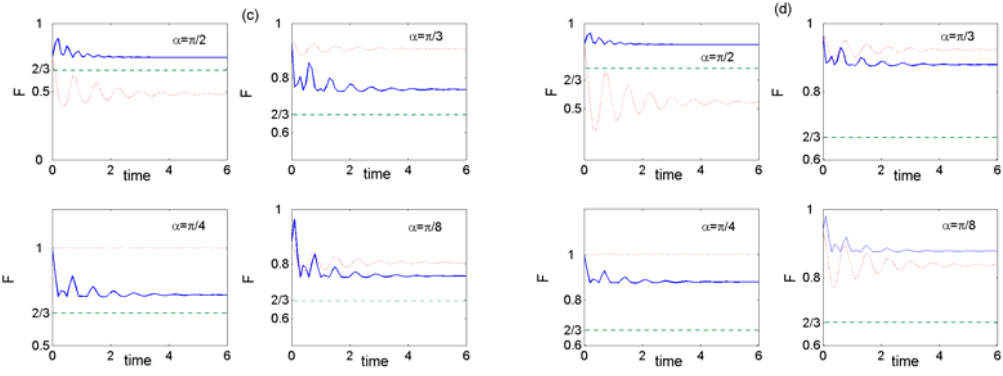

**Fig. 9** Dynamics of fidelity for $D_z = 2$ (blue solid line) and $D_x = 2$ (red dotted line) when the initial state is $|\varphi(0)\rangle = \cos\alpha|01\rangle + \sin\alpha|10\rangle$. $J = 1$, $\gamma = 0.4$, $J_z = 0.5$, $\Gamma = 0.02$, $\phi = 0$. (a) $\theta = \pi/2$, (b) $\theta = \pi/3$, (c) $\theta = \pi/4$, (d) $\theta = \pi/6$.

In Fig. 9, fidelity $F$ is plotted as a function of the time $t$ for different initial state $\alpha$ and inputting state $\theta$. For the purpose of transmit $\rho_{in}$ with better fidelity than any classical communication protocol, we require Eq. (13) to be strictly greater than 2/3[24]. Figure 9a gives the evolution of fidelity for the $D_{z/x}$ system when the inputting state is $|\psi_{in}\rangle = \sqrt{2}/2|10\rangle + \sqrt{2}/2|01\rangle$. In this figure, the $D_z$ system all behaves inferior to the classical communication except for some small region. But $D_x$ system performs better in this condition except for initial state $|\varphi(0)\rangle = |10\rangle$. If we select the inputting state $|\psi_{in}\rangle = \sqrt{3}/2|10\rangle + 1/2|01\rangle$, the result in Fig. 9b shows that the two kinds of system all behave better than the cases before. The entanglement will become a stability constant with the decoherence time and this is meaningful for quantum information processing. In Fig. 9a, Fig. 9b, and Fig. 9c, we find the $D_x$ system is always superior to the $D_z$ system except the initial state $|\varphi(0)\rangle = |10\rangle$. The thing changes for Fig. 9d, the inputting state is $|\psi_{in}\rangle = 0.9659|10\rangle + 0.2588|01\rangle$ in this case, if we choose the initial state $|\varphi(0)\rangle = 0.9238|01\rangle + 0.3826|10\rangle$, the $D_z$ system achieves better fidelity than the $D_x$ system. This indicates that seemingly some system are not suitable for

teleportation, they can acquire some best exhibition if we take proper system parameters and the fidelity can be larger than 2/3. Moreover, we find that with the decrease of the inputting state angel $\theta$, the fidelity of two kinds of system behave better than before.

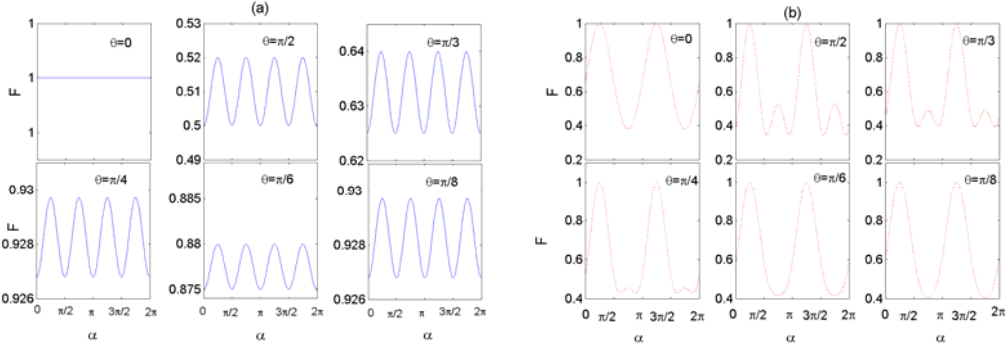

**Fig. 10** Asymptotical behavior of the fidelity versus the initial state angle $\alpha$ when the initial state in $|\varphi(0)\rangle = \cos\alpha |01\rangle + \sin\alpha |10\rangle$, $J = 1$, $\gamma = 0.8$, $J_z = 2$, $\Gamma = 0.02$, $\phi = 0$. (a) $D_z = 2$, (b) $D_x = 2$.

The asymptotical behavior of the fidelity versus the model parameters $\alpha$ for initial state $|\varphi(0)\rangle = \cos\alpha |01\rangle + \sin\alpha |10\rangle$ is shown in the Fig. 10. Except the case of $\theta = 0$ in Fig. 10(a), the asymptotic fidelity $F$ also demonstrates periodicity. This means we can implement teleportation process in proper angle and get optimal fidelity. We note that the cycle is $\pi/2$, $\pi$ for $D_z$ and $D_x$ system respectively. Interestingly, we can get the ideal fidelity if the inputting state $\theta$ gets some small values for the $D_{z/x}$ system. These result also indicates that the output entanglement can not control the asymptotical behavior of the fidelity. Comparing the two figures, we find that the $D_z$ system behaves better than $D_x$ system in quantum teleportation for $\theta < \pi/4$, while on $\theta > \pi/4$ the $D_x$ system gets the run upon.

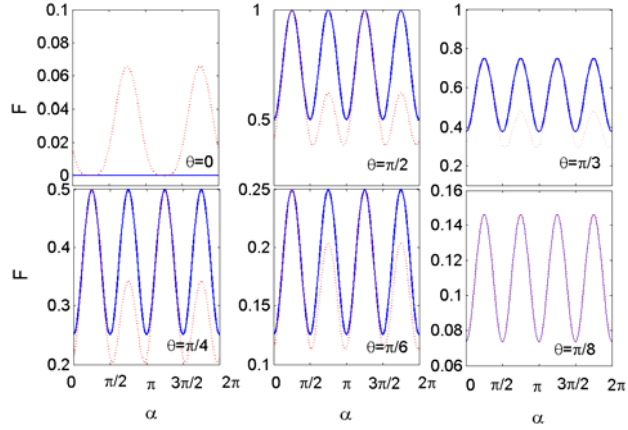

**Fig. 11** Asymptotical behavior of the fidelity versus the initial state angle $\alpha$ for $D_z = 2$ (blue solid line) and $D_x = 2$ (red dotted line) when the initial state in $|\varphi(0)\rangle = \cos\alpha |00\rangle + \sin\alpha |11\rangle$, $J = 1$, $\gamma = 0.1 = 0.1$, $J_z = 3$, $\Gamma = 0.02$, $\varphi = 0$.

In Fig. 11 we also give the behaviors of asymptotical fidelity for another kind of initial state $|\varphi(0)\rangle = \cos\alpha |00\rangle + \sin\alpha |11\rangle$. Obviously, this initial state does not suit for entanglement teleportation in the majority. But in some circumstance, this initial state can be chosen for transmit unknown state, i.e. $\theta = \pi/2, \theta = \pi/3$. We also find the $D_z$ system exceeds the $D_x$ system for quantum teleportation in this case.

# 6 Summary

In summary, we investigate the two qubit Heisenberg XYZ model with different Dzyaloshinskii-Moriya interaction and calculate the entanglement teleportation, the fidelity of entanglement teleportation with the Milburn's intrinsic decoherence taken into consideration. The results demonstrate that after reaching the maximum decoherence time the entanglement will not decrease but get a steady value for some initial state. The maximum values of the asymptotic fidelity $F$ exhibit periodic. Moreover, the result show that the $D_z$ system behaves better than $D_x$ system in quantum teleportation for $\theta < \pi/4$, while for the other cases the $D_x$ system gets the run upon. These results are valuable in quantum information processing based on the solid state qubits. Also, it may be helpful to recognize how the decoherence factors affect the entanglement teleportation.

Acknowledgments

This work was supported by the National Natural Science Foundation of China (grants nos 11035001, 10735010, 10975072, 11375086 and 11120101005), by the 973 National Major State Basic Research and Development of China (grants nos 2010CB327803 and 2013CB834400), by CASKnowledge Innovation project no KJCX2-SW-N02, by Research Fund of Doctoral Point (RFDP) grant no 20100091110028, by the Research and Innovation Project for College Postgraduate of JiangSu Province grant nos CXLX13-025 and CXZZ13- 0032, by the National Natural Science Foundation of China (grants no 11347112), by the Project Funded by the Priority Academic Program Development of Jiangsu Higher Education Institutions (PAPD) and by the Deutsche Forschungsgemeinschaft through TRR80.